\documentstyle[aps,prl,twocolumn,epsf]{revtex}

\begin{document}

\hyphenation{Sr}

\twocolumn[
\hsize\textwidth\columnwidth\hsize\csname@twocolumnfalse\endcsname
\draft

\title{Non-Universal Power Law of the \lq \lq Hall Scattering Rate" 
in a Single-Layer Cuprate Bi$_{2}$Sr$_{2-x}$La$_x$CuO$_{6}$}

\author{Yoichi Ando$^{1,2}$ and T. Murayama$^{1,2}$}
\address{$^1$Central Research Institute of Electric Power
Industry, Komae, Tokyo 201-8511, Japan}
\address{$^2$Department of Physics, Science University of Tokyo,
Shinjuku-ku, Tokyo 162-8601, Japan}

\date{Received BSCO-3.tex}
\maketitle

\begin{abstract}
In-plane resistivity, Hall coefficient, and magnetoresistance (MR) are
measured in a series of high-quality Bi$_{2}$Sr$_{2-x}$La$_x$CuO$_{6}$
crystals with various carrier concentrations, from underdope
to overdope.  It is found that the temperature dependence of the 
Hall angle obeys a power law $T^{\alpha}$ with $\alpha$ 
systematically decreasing with increasing doping, which questions the
universality of the Fermi-liquid-like $T^2$ dependence of the 
\lq \lq Hall scattering rate".
The systematics of MR indicates an increasing role of spin scattering
in underdoped samples.
\end{abstract}

\pacs{PACS numbers: 74.25.Fy, 74.62.Dh, 74.72.Hs}
]

The peculiar normal-state properties of high-$T_c$ cuprates
are generally considered to be the keys to elucidate the
high-$T_c$ mechanism, since they give us a clue to clarify
the nature of the strongly-correlated electronic state of
the cuprates.
To understand the underlying electronic state of the cuprates, 
it is desirable to study the normal state at low temperatures 
\cite{logT}
and to chase the systematic evolution thereof with carrier concentration
\cite{Boebinger}.
Therefore, particularly useful system for the normal-state study 
is such a system where
$T_c$ is relatively low and where the the carrier concentration 
can be changed in a wide range from underdope to overdope.

A typical study in which a wide temperature window for
the normal-state is desirable is the measurement of the 
temperature dependence of the scattering time.
For example, the observation of the $T$-linear resistivity from
10 K to 700 K in Bi$_{2}$Sr$_{2}$CuO$_{6}$ (Bi-2201) \cite{Martin}
had a strong impact, because it clearly demonstrated
the dominance of the electron-electron interaction in the
scattering mechanism.
In high-$T_c$ cuprates, it has been discussed that the charge transport
is governed by two different scattering times with different 
temperature dependences \cite{Chien,Anderson,Coleman}; 
according to this 
\lq \lq two-scattering-time" model, in-plane resistivity
$\rho_{ab}$ is governed by the 
transport scattering time $\tau_{tr}$ ($\sim$$T^{-1}$) 
and the Hall angle $\theta_H$ is governed by the 
\lq \lq Hall scattering time" $\tau_{H}$ ($\sim$$T^{-2}$).
However, this idea of \lq \lq scattering-time separation" has
not yet gained a complete consensus and
there are other approaches to understand the unusual 
normal-state transport properties \cite{Stojkovic,Ioffe}.
Therefore, it would be useful to establish the temperature and 
doping range in which the $T^2$ behavior of the 
Hall scattering rate $\tau_{H}^{-1}$ is observed.

There are two widely-known cuprate systems which satisfy the requirements
of the relatively low $T_c$ and the availability of a wide doping range; 
La$_{2-x}$Sr$_{x}$CuO$_4$ (LSCO) system and Bi-2201 system.
Bi-2201 has not been as intensively studied as LSCO,
mostly because of the difficulty in obtaining high-quality
single crystals.  A number of problems have been known for
Bi-2201 crystals: 
(a) the transport properties of B-2201
are quite non-reproducible even among crystals of nominally the 
same composition \cite{Jin,Mackenzie,Ando};
(b) the residual resistivity of the in-plane resistivity $\rho_{ab}$
is usually large (the smallest value reported to date is 
70 $\mu \Omega$cm \cite{Martin,Ando}), as opposed to LSCO, 
YBa$_{2}$Cu$_3$O$_7$ (YBCO), or Bi$_{2}$Sr$_{2}$CaCu$_2$O$_{8}$  
(Bi-2212), where the residual resistivity in high-quality
crystals is negligibly small; and
(c) the temperature dependence of the Hall coefficient is weak and 
thus the Hall angle $\theta_H$ does not obey the $T^2$ law 
\cite{Mackenzie,Ando}, while $\theta_H \sim T^2$ has been
almost universally observed in other cuprates \cite{Coleman}.
On the other hand, the Bi-2201 system has very attractive 
characteristics: it has been known for Bi-2201 
that the carrier concentration can be widely changed by partially
replacing Sr with La (to underdope) or Bi with Pb (to overdope)
\cite{Maeda}; at optimum doping 
(Bi$_{2}$Sr$_{2-x}$La$_x$CuO$_{6}$ with $x$$\simeq$0.4),
the maximum $T_c$ is about 30 K \cite{Maeda,Yoshizaki}, which
is lower than the maximum $T_c$ of LSCO. 
Therefore, if single crystals of sufficiently high quality are
grown, this system would present an ideal stage for the systematic 
study of the normal-state properties down to lower temperatures
than in other cuprates.

In this Letter, we report that it is possible to obtain a series 
of high-quality Bi-2201 crystals and show that in those high-quality 
crystals the normal-state transport properties display behaviors
which are in good accord with other cuprates; 
for example, in the underdoped region, 
$\rho_{ab}$ shows a downward deviation from the $T$-linear 
behavior at a certain temperature which decreases with increasing
doping and, in the overdoped region, $T$ dependence of $\rho_{ab}$
changes to $T^n$ with $n>$1.  
What is new in this system
is that the Hall angle indeed obeys a power law $T^{\alpha}$ with 
$\alpha$$\simeq$2, but the power $\alpha$ shows a systematic 
decrease towards smaller
values as the carrier concentration is increased.
This finding of a systematic change of $\alpha$ in a relatively-simple
single-layer cuprate system poses a serious question to the universality
of the \lq \lq Fermi-liquid like" $T^2$ behavior of the Hall
scattering rate $\tau_H^{-1}$ \cite{Coleman}. 
Another notable finding is that
the ratio of the longitudinal magnetoresistance (MR) to the transverse
MR increases systematically with decreasing carrier concentration, 
indicating that a spin contribution to the MR gradually 
becomes significant as the sample becomes more underdoped.

The single crystals of Bi$_{2}$Sr$_{2-x}$La$_x$CuO$_{6}$ (BSLCO) 
are grown using a floating-zone technique in 1 atmosphere of 
flowing oxygen.
It is known that pure Bi-2201 is an overdoped system \cite{Maeda}.
Since the La substitution to the Sr site reduces the number of holes, 
increasing La doping brings the system from overdoped region 
to underdoped region.
We found that the La substitution results in a growth of crystals 
of better morphology compared to the pure Bi-2201.
The actual La concentrations in the crystals are determined by the 
inductively-coupled plasma spectrometry (ICP) technique.
Here we report crystals with $x$=0.24, 0.30, 0.44, 0.57, and 0.66, 
for which the zero-resistance $T_c$ is 24, 30, 33, 28.5, and 20 K,
respectively.  The inset to Fig. 1 shows the zero-resistance $T_c$
(and the mid-point $T_c$) as a function of $x$. 
Apparently, $x$$\simeq$0.4 corresponds to the 
optimum doping, which is consistent with previous reports on BSLCO
\cite{Maeda,Yoshizaki}.
The onset of the Meissner effect (measured with a commercially available
Quantum Design SQUID magnetometer) for the optimally-doped crystals 
is 33 K.  To our knowledge, this optimum $T_c$ is the highest
value ever reported for Bi-2201 or BSLCO system.
The ICP analysis found that our crystals are Bi-deficient by about 0.2,
while Sr+La is almost stoichiometric;
for example, the $x$=0.44 sample has the composition of
Bi$_{1.8}$Sr$_{1.57}$La$_{0.44}$CuO$_{6+\delta}$.

\vspace{-0.7cm}
\begin{figure}[htbp]
\begin{center}
 \epsfxsize=75mm
 $$\epsffile{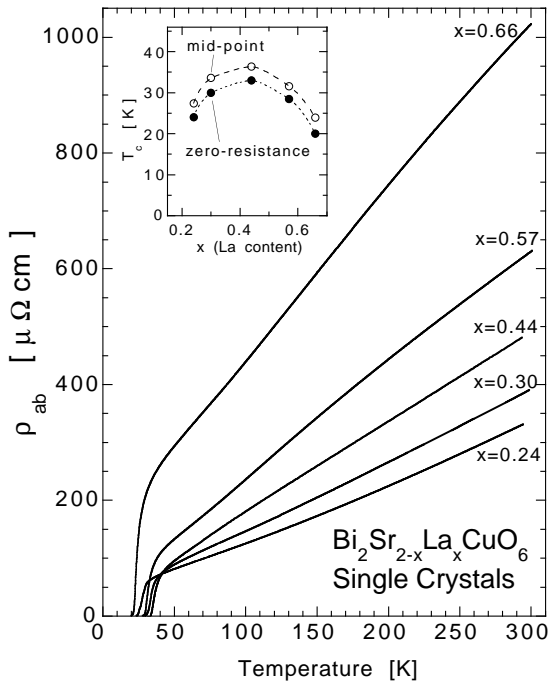}$$
\end{center}
\caption{$T$ dependence of $\rho_{ab}$ of the BSLCO crystals.
Inset: $T_c$ as a function of La content $x$.  The dashed lines are
guide to the eyes.}
\label{fig1}
\end{figure}

The crystals are cut into a rectangular shape with typical size of 
2 $\times$ 1 $\times$ 0.015 mm$^3$.
The thickness of the crystals are accurately determined 
by measuring the weight of the sample with 0.1 $\mu$g resolution;
therefore, the uncertainty in determining the magnitude of the
resistivity is less than $\pm$5\%.
The crystals are annealed at 400$^{\circ}$C for 30 minutes in flowing
oxygen upon firing silver epoxy.
We use a standard six-terminal method for simultaneous
magnetoresistance (MR) and $R_H$ measurements, in which
the data are taken with an ac technique in the sweeping 
magnetic field at fixed temperatures.
The temperature is very carefully controlled and stabilized using 
both a capacitance sensor and a Cernox resistance sensor
to avoid systematic temperature deviations with magnetic fields.
The stability of the temperature during the MR and $R_H$
measurements is within a few mK.

Figure 1 shows the temperature dependence of $\rho_{ab}$ for the
five $x$ values in zero field.
Clearly, both the magnitude of $\rho_{ab}$ and its slope  
show a systematic decrease with increasing carrier concentration
(decreasing $x$).
We found that it is only at the optimum doping that $\rho_{ab}$
shows a perfect $T$-linear behavior.  Figure 2 shows the
temperature dependence of the slope $d\rho_{ab}/dT$;
only the $x$=0.44 sample shows a constant $d\rho_{ab}/dT$,
which corresponds to the $T$-linear behavior, in a
wide temperature range (from 300 K to 120 K).
We note that a fitting of the $\rho_{ab}$ data of 
the $x$=0.44 sample to $\rho_{ab}$=$\rho_0+AT$ gives the
residual resistivity $\rho_0$ of only 25 $\mu \Omega$cm,
which, to our knowledge, is the smallest value ever reported for
pure Bi-2201 or BSLCO.

\vspace{-0.7cm}
\begin{figure}[htbp]
\begin{center}
 \epsfxsize=75mm
 $$\epsffile{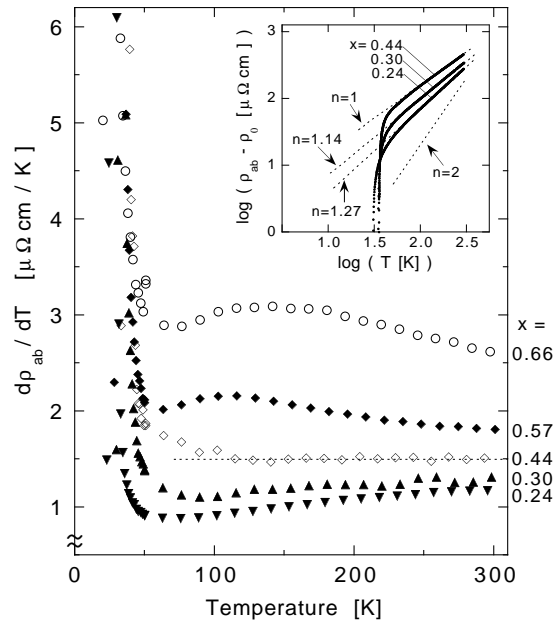}$$
\end{center}
\caption{$T$ dependence of the slope $d\rho_{ab}/dT$.
The dashed line represents a constant slope ($T$-linear behavior).
Inset: Plot of $\log (\rho_{ab}$-$\rho_0)$ vs $\log T$ for three
$x$ values to show the power-law dependence $\rho_{ab}$=$\rho_0+AT^n$,
which is represented by dotted lines.}
\label{fig2}
\end{figure}

The underdoped samples ($x$=0.57 and 0.66) show a rather
complex temperature dependence in $\rho_{ab}$, which is 
similar to that of underdoped YBCO \cite{Ito};
the behavior in underdoped YBCO is characterized by a downward
deviation from the $T$-linear dependence and the presence of a
maximum in $d\rho_{ab}/dT$.  Such a behavior has been correlated
to the opening of a pseudogap \cite{Ito}.
The broad maximum in $d\rho_{ab}/dT$ moves to higher
temperature as the carrier concentration is decreased,
which agrees with the conjecture that the pseudogap opens at higher
temperature in more underdoped samples.
On the other hand, the slope $d\rho_{ab}/dT$ of the overdoped samples 
($x$=0.30 and 0.24) monotonically decreases with decreasing temperature. 
This reflects the fact that $\rho_{ab}$ of
the overdoped samples behaves as $\rho_{ab}$=$\rho_0+AT^n$ 
with $n$ larger than 1, 
a behavior reported in the overdoped samples of LSCO \cite{Takagi}
and Tl$_{2}$Ba$_{2}$CuO$_{6+\delta}$ (Tl-2201) \cite{Kubo}.  
The inset to Fig. 2 shows that $\rho_{ab}(T)$ for $x$=0.30 and 0.24
can actually be described by the power law with $n$=1.14 and 1.27, 
respectively. 
Therefore, in both the underdoped and overdoped regions,
the behavior of $\rho_{ab}(T)$ shows an evolution 
which can be considered to be \lq \lq standard" for high-$T_c$ cuprates.
This observation indicates that Bi-2201 is not an exceptional system 
but rather is a promising system for the systematic study of the 
normal-state transport properties.

\vspace{-0.7cm}
\begin{figure}[htbp]
\begin{center}
 \epsfxsize=75mm
 $$\epsffile{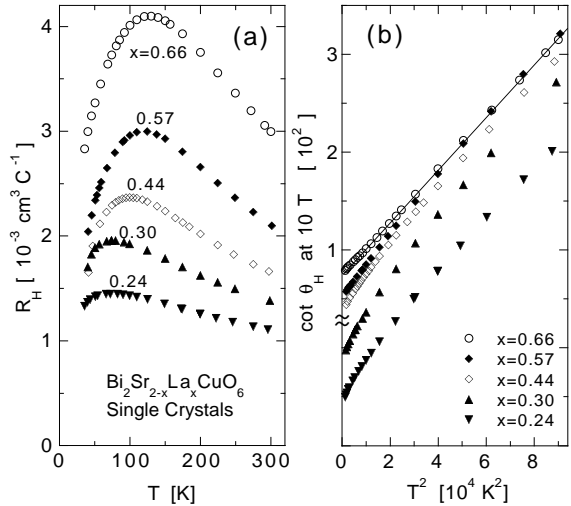}$$
\end{center}
\caption{(a) $T$ dependence of $R_H$ of the BSLCO crystals.
(b) Plot of $\cot \theta_H$ vs $T^2$ (the data for $x$=0.30 and 0.24
are shifted down by 50 and 100, respectively, to avoid congestion).  
The solid line is a fit to the $x$=0.66 data.}
\label{fig3}
\end{figure}

Figure 3(a) shows the temperature dependence of $R_H$ for the five samples.
A clear evolution of $R_H$ with $x$ is observed; the change in the 
magnitude of $R_H$ at 300 K suggests that the carrier concentration is
actually reduced roughly by a factor of 3 upon increasing $x$ 
from 0.24 to 0.66.
Figure 3(b) shows the plot of $\cot \theta_H$ (=$\rho_{xx}$/$\rho_{xy}$) 
vs $T^2$.  Only the data for $x$=0.66 can be fitted with a straight
line in this plot, indicating that the $T^2$ law of $\cot \theta_H$
holds only in this most underdoped sample.
We found that $\cot \theta_H$ for other $x$ values obey a 
power law $T^{\alpha}$ with $\alpha$ smaller than 2, which is shown in
Figs. 4 (a) - (d).  The best powers are 1.85, 1.70, 1.65, and 1.60,
for $x$ =0.57, 0.44, 0.30, and 0.24, respectively.  In all the panels
of Fig. 4, the data are very well fitted with straight lines.
Therefore, the power-law temperature dependence of
the Hall scattering rate $\tau_H^{-1}$ holds for every doping in 
BSLCO, but the power $\alpha$ shows a systematic decrease with
increasing carrier concentration \cite{footnote}.  
A particularly intriguing fact here is that $\cot \theta_H$ of the 
optimally-doped sample changes as $T^{1.70}$, not as $T^2$,
while $\rho_{ab}$ shows a good $T$-linear behavior.
This might mean that the \lq \lq Fermi-liquid like" behavior of 
$\tau_H^{-1}$$\sim$$T^2$ (which has been proposed to be the 
characteristic of spinons \cite{Anderson}) 
may not be a generic feature of the optimally-doped cuprates.

\vspace{-0.7cm}
\begin{figure}[htbp]
\begin{center}
 \epsfxsize=75mm
 $$\epsffile{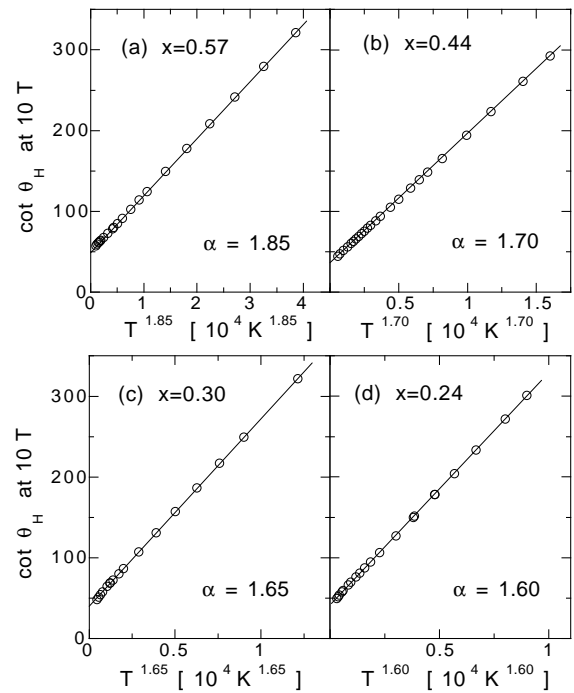}$$
\end{center}
\caption{Plots of $\cot \theta_H$ vs $T^{\alpha}$ for the data of
(a) $x$=0.57, (b) $x$=0.44, (c) $x$=0.30, and (d) $x$=0.24.
The solid lines are fits to the data.}
\label{fig4}
\end{figure}

Figure 5 shows the transverse and longitudinal MR of four of the samples,
$x$=0.24, 0.44, 0.57, and 0.66.  
One may immediately notice a trend that the 
relative magnitude of the longitudinal MR compared to the transverse MR 
increases with increasing $x$.
(We omit the result of $x$=0.30 due to the limited space; 
the behavior of this sample fits well into the trend.)
Since it is expected that the longitudinal MR mostly comes from a
spin contribution (while the transverse MR consists of both
an orbital and the spin contribution), our result suggests that
the role of spins in the transport becomes increasingly significant 
as the sample becomes more underdoped.

\vspace{-0.7cm}
\begin{figure}[htbp]
\begin{center}
 \epsfxsize=75mm
 $$\epsffile{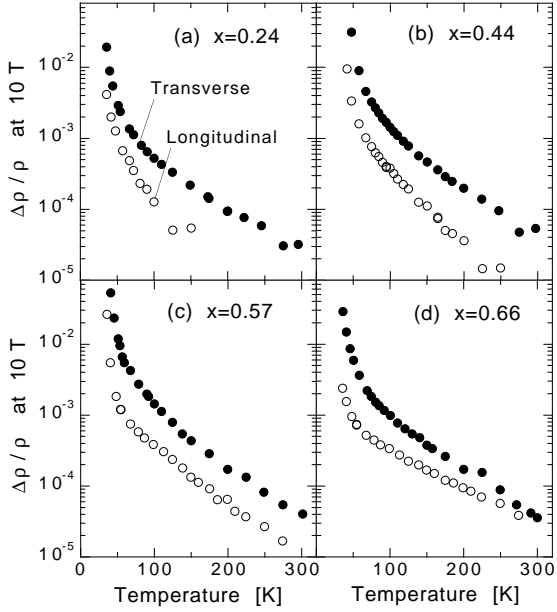}$$
\end{center}
\caption{Transverse MR (solid circles) and 
longitudinal MR (open circles) for
(a) $x$=0.24, (b) $x$=0.44, (c) $x$=0.57, and (d) $x$=0.66.}
\label{fig5}
\end{figure}

In passing, let us briefly mention our analysis to see if any 
scaling is applicable to the MR data.
Firstly, we found that the classical Kohler's rule 
$\Delta \rho/\rho$$\sim$$(H/\rho)^2$ is strongly violated, 
as in other cuprates \cite{Kimura,Harris}.
We also found that the \lq \lq modified Kohler's rule"
$\Delta \rho/\rho$$\sim$$(\cot \theta_H)^{-2}$ 
is not very well applicable to our data; this is different from
the results for LSCO and YBCO \cite{Kimura,Harris} and 
might be related to the fact that $\cot \theta_H$ does not behave as 
$T^2$ except for the $x$=0.66 sample.
The details of the MR analysis will be published elsewhere.

The above results tell us altogether that the systematics of the
power law of the scattering rates ($\tau_{tr}^{-1}$$\sim$$T^{n}$ 
and $\tau_H^{-1}$$\sim$$T^{\alpha}$)
needs to be reconsidered.
In particular, $\tau_H^{-1}$ shows a tendency that the power 
$\alpha$ becomes systematically smaller with increasing doping 
in the whole doping range studied.
This observation, combined with the systematic change of the 
longitudinal MR, might suggest that the $T^2$ law of 
$\tau_H^{-1}$ is observable only when the spin contribution to
the charge transport is strong.  This in turn suggests that in Bi-2201
the role of spin degrees of freedom is already weakened at optimum
doping, which might be the reason for the relatively low $T_c$ of
this system.  Interestingly, another single-layer
cuprate Tl-2201 shows a good $T^2$ dependence of $\tau_H^{-1}$ 
at optimum doping \cite{Andre} 
and Tl-2201 has the maximum $T_c$ of 85 K.
Finally, because of the relatively wide temperature range in which the
normal-state transport properties can be studied,
BSLCO system offers an ideal stage for the detailed study of the
systematic evolution of the scattering times as well as other 
normal-state properties.

We thank A.N. Lavrov, N. Nagaosa and A.J. Schofield for helpful discussions. 

%


\begin{references}
\bibitem{footnote}
A similar trend has been reported in overdoped LSCO in a much narrower
doping range \cite{Kimura}.
\end{references}
\end{document}